\documentclass{PoS}

\usepackage{graphicx}
\usepackage{amsthm,amsmath,amssymb,bm}

\title{Lattice QCD simulation of the Berry curvature}

\ShortTitle{Lattice QCD simulation of the Berry curvature}

\author{Arata~Yamamoto\\
Department of Physics, The University of Tokyo, Tokyo 113-0033, Japan\\
E-mail: \email{arayamamoto@nt.phys.s.u-tokyo.ac.jp}}

\abstract{
The Berry curvature is a fundamental concept describing topological order of quantum systems.
While it can be analytically tractable in non-interacting systems, numerical simulations are necessary in interacting systems.
We present a formulation to calculate the Berry curvature in lattice QCD.
}

\FullConference{34th annual International Symposium on Lattice Field Theory\\
                 24-30 July  2016\\
                 University of Southampton, Southampton, UK}

\begin{document}

\section{Berry curvature}

Let us consider the Hamiltonian $H(p)$ described by a parameter $p$.
From the eigenvalue equation
\begin{equation}
H(p) \Phi(p) = E(p) \Phi(p)
,
\end{equation}
the eigenfunction $\Phi(p)$ is obtained.
The Berry connection is defined as
\begin{equation}
\tilde{A}_\mu(p) = -i \Phi^\dagger(p) \frac{\partial}{\partial p^\mu} \Phi(p)
\end{equation}
and the Berry curvature is defined as
\begin{equation}
\tilde{F}_{\mu\nu}(p) = \frac{\partial}{\partial p^\mu} \tilde{A}_\nu(p) - \frac{\partial}{\partial p^\nu} \tilde{A}_\mu(p)
.
\end{equation}
These definitions indicate that the Berry connection and curvature correspond to the gauge connection and curvature in parameter space, respectively.

The Berry curvature is quite general in theoretical physics \cite{Berry:1984}.
It can be defined in any parameter space.
In this study, we focus on the spatial momenta of the ground-state fermions.
The Berry curvature of fermions is essential for describing several physical phenomena.
For example, the Berry curvature of chiral fermions describes the chiral magnetic and vortical effects \cite{Son:2012wh}, and the Berry curvature of electrons describes the quantum Hall effect \cite{Thouless:1982zz} and topological insulators \cite{Kane:2005}.
Although the Berry curvature has been calculated in many theoretical works, most of the calculations were done in non-interacting approximation.
For the Berry curvature including interaction effects, we need numerical simulations.

For this purpose, we formulate the computational scheme to calculate the Berry curvature in lattice QCD.
This presentation is based on the recent paper \cite{Yamamoto:2016rfr}.
We would like to skip technical details and overview only the outline.

\section{Formalism}

To calculate the Berry curvature, we need the fermion ground state $\Phi(p)$ as a function of the spatial momentum $p$.
The ground state is obtained by the standard ground-state projection in lattice QCD, which is used for the ground-state hadron mass calculation.
We construct a single-fermion state with a fixed spatial momentum by the spatial Fourier transformation
\begin{equation}
\phi (p,\tau) = \sum_{x,x'} e^{ip\cdot(x-x')} D^{-1}(x,\tau|x',0) \phi_{\rm init} 
,
\end{equation}
where $D^{-1}(x,\tau|x',\tau')$ is a fermion propagator and $\phi_{\rm init}$ is an initial state.
When the imaginary-time separation $\tau$ is large enough, this state is independent of the choice for $\phi_{\rm init}$ and goes to the ground state
\begin{equation}
\Phi(p) = \lim_{\tau \to \infty} \phi (p,\tau)
.
\end{equation}
We here consider a non-degenerate ground state for simplicity.
When $\Phi(p)$ is not degenerate, the Berry curvature is Abelian.
In general, the Berry curvature can be defined for any state.
For example, degenerate states have the non-Abelian Berry curvature \cite{Wilczek:1984dh}.
The formulation will be easily extended to the non-Abelian case.

\begin{figure}[h]
\centering
\includegraphics[scale=0.7]{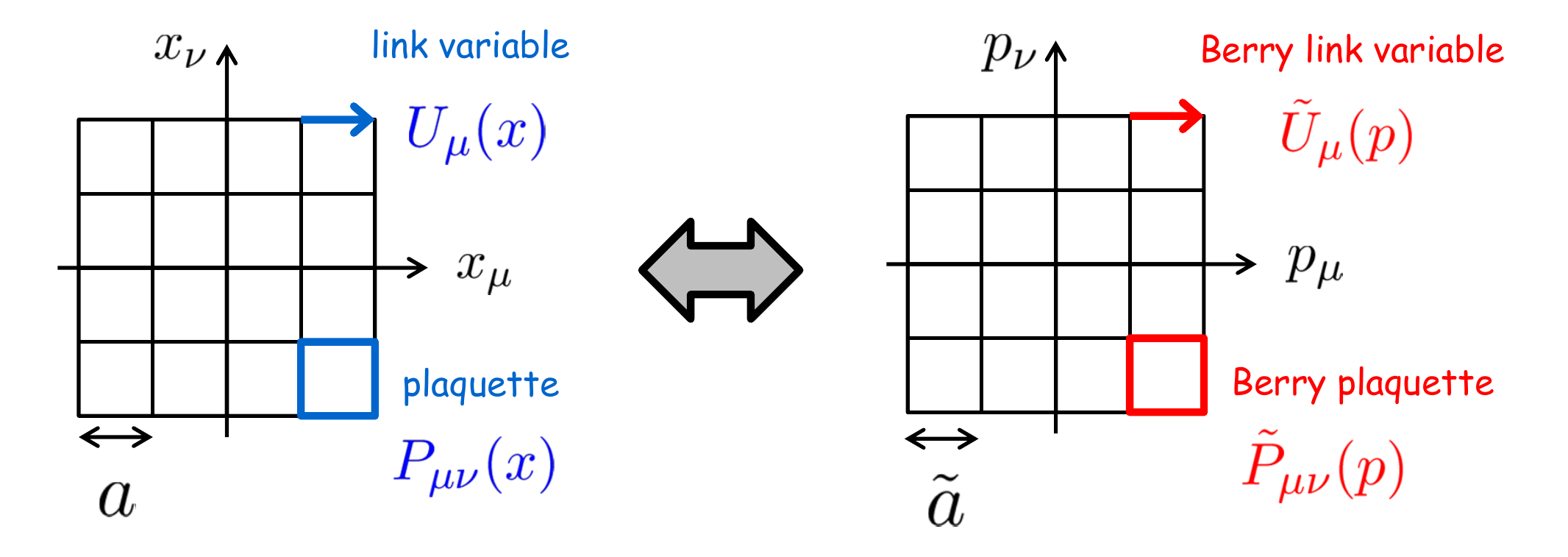}
\caption{\label{figlattice}
The schematic figure of the standard lattice gauge theory (left) and the lattice Berry field theory (right).
}
\end{figure}

Since lattice QCD simulations are done in a finite volume, the corresponding momentum space is also a finite-volume lattice.
Thus we should formulate the Berry curvature as lattice gauge theory in momentum space \cite{Fukui:2005wr}.
The schematic figure is shown in Fig.~\ref{figlattice}.
The coordinate-space lattice with the spacing $a$ and the length $L$ is mapped onto the momentum-space lattice with the spacing $\tilde{a}=2\pi/L$ and the length $\tilde{L}=2\pi/a$.
Since gauge connection is introduced as link variable, the Berry connection is introduced as the Berry link variable
\begin{equation}
\tilde{U}_\mu(p) = e^{i\tilde{a} \tilde{A}_\mu(p)} = \frac{\Phi^\dagger(p) \Phi(p+\tilde{\mu})}{|\Phi^\dagger(p) \Phi(p+\tilde{\mu})|}
,
\end{equation}
where $\tilde{\mu}$ is the unit vector in the $\mu$ direction on the momentum-space lattice.
The Berry curvature is given by the Berry plaquette
\begin{equation}
\tilde{P}_{\mu\nu}(p) = e^{i\tilde{a}^2 \tilde{F}_{\mu\nu}(p)} = \tilde{U}_\mu(p) \tilde{U}_\nu(p+\tilde{\mu}) \tilde{U}^*_\mu(p+\tilde{\nu}) \tilde{U}^*_\nu(p)
.
\end{equation}
In the Monte Carlo simulation, we calculate this Berry curvature for each configuration, and take the ensemble average over configurations.

The Berry connection has the local U(1) gauge degree of freedom, which exists even in a non-interacting case.
The Berry curvature is independent of the U(1) gauge choice because the plaquette is gauge invariant in lattice gauge theory.
In an interacting case, there is additional gauge, i.e., the local SU(3) gauge of gluons.
Since a single fermion is gauge dependent, the Berry curvature depends on the SU(3) gauge choice.
Thus we did not fix the U(1) gauge but fixed the SU(3) gauge in the simulation below.

\section{Example}

We performed the first numerical test in a simple example.
We considered the (2+1)-dimensional Wilson fermion
\begin{equation}
 D(x,x') 
= (ma+3) \delta_{x,x'}- \frac{1}{2} \sum_{\mu=1}^{3} \bigg[ \left(1-\sigma_\mu \right) U_\mu(x) \delta_{x+\hat{\mu},x'}
+ \left(1+\sigma_\mu \right) U_\mu^\dagger(x') \delta_{x-\hat{\mu},x'} \bigg]
,
\end{equation}
where $\hat{\mu}$ is the unit vector in the $\mu$ direction on the coordinate-space lattice.
In 2+1 dimensions ,the Berry link variable is described by two-dimensional U(1) lattice gauge theory.
The analysis is the same as the two-dimensional U(1) lattice gauge theory \cite{Panagiotakopoulos}.
We calculated the Berry curvature
\begin{equation}
\tilde{a}^2 \tilde{F}_{xy}(p) = {\rm Im \ ln} \ \tilde{P}_{xy}(p)
.
\end{equation}
This corresponds to the topological charge density in the two-dimensional U(1) lattice gauge theory.
The integral of the topological charge density gives topological charge
\begin{equation}
N = \frac{1}{2\pi} \sum_p \tilde{a}^2 \tilde{F}_{xy}(p)
,
\end{equation}
which is called the first Chern number.
The first Chern number is topological charge and thus an integer.

\begin{figure}[h]
\centering
\includegraphics[scale=0.5]{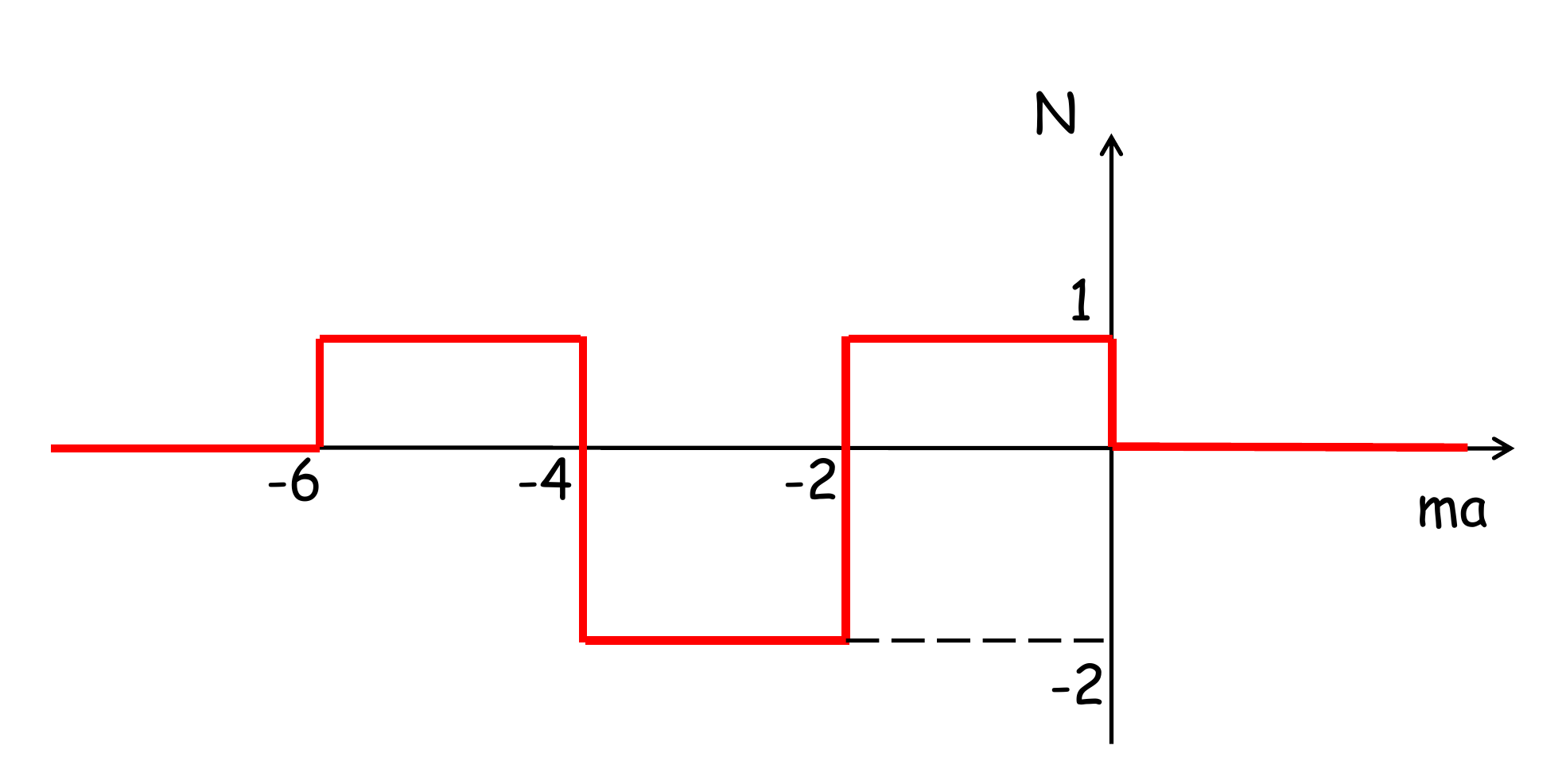}
\caption{\label{figfree}
The first Chern number $N$ of the free (2+1)-dimensional Wilson fermion as a function of the bare mass parameter $m$.
}
\end{figure}

The first Chern number of the non-interacting Wilson fermion is known.
It is depicted in Fig.~\ref{figfree}.
For positive mass $m>0$, topology is trivial, i.e., $N=0$.
On the other hand, for negative mass $m<0$, topology can be nontrivial, i.e., $N \ne 0$.
This behavior can be easily understood by counting the numbers of massless modes.
The Chern number changes when massless modes appear.
The (2+1)-dimensional Wilson fermion has one physical mode and seven doublers.
The physical mode is massless at $m=0$.
This gives the change from $N=1$ to $N=0$ at $m=0$.
The doublers are massless only in $m<0$.
These give the changes in $m<0$.
In particle physics, these doubler poles are unphysical because particle mass must be $m>0$.
In condensed matter physics, however, materials with $m<0$ can be generated.
Actually, the non-relativistic version of the Wilson fermion is used for a model of the quantum Hall effect \cite{Qi:2006}.
The first Chern number explains the quantization of the Hall resistivity $R_{xy} = 2\pi/e^2 N$ \cite{Thouless:1982zz}.

\begin{figure}[h]
\centering
\includegraphics[scale=0.85]{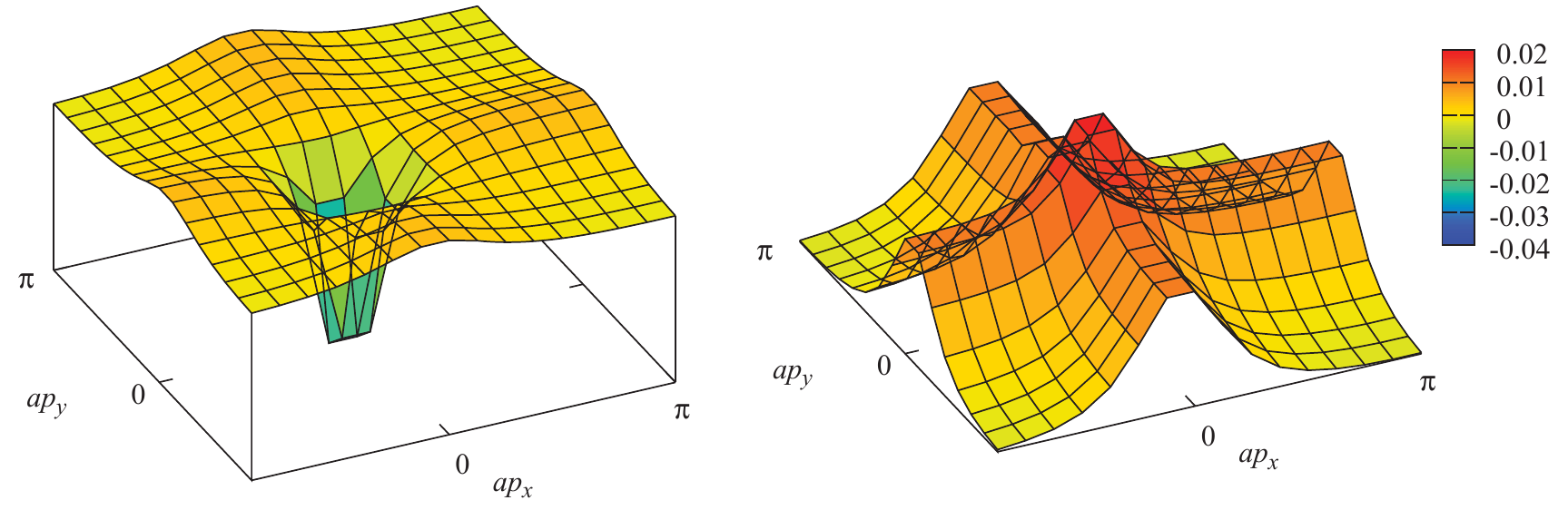}
\caption{\label{figBC}
The Berry curvature $\tilde{a}^2 \tilde{F}_{xy}(p)$ at $ma=0.5$ (left) and $ma=-0.5$ (right).
}
\end{figure}

\begin{figure}[h]
\centering
\includegraphics[scale=0.75]{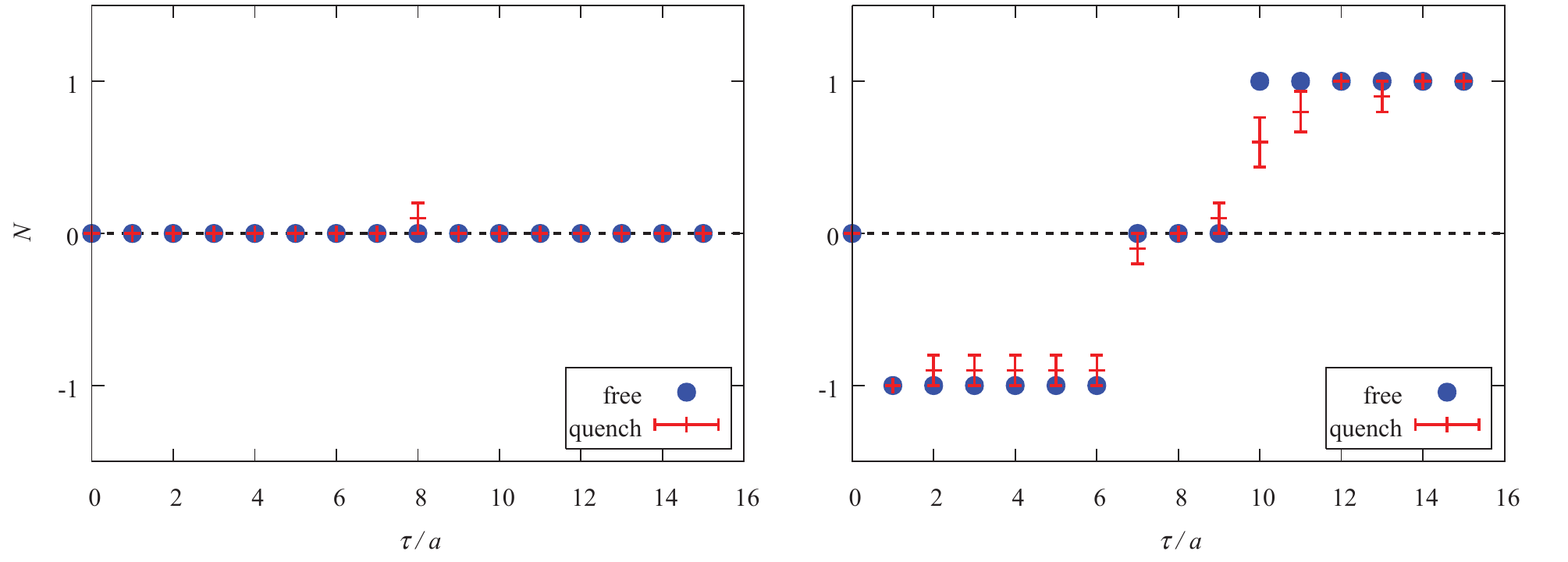}
\caption{\label{figN}
The first Chern number $N$ as a function of imaginary time $\tau$ at $ma=0.5$ (left) and $ma=-0.5$ (right).
}
\end{figure}

In Fig.~\ref{figBC}, the momentum-space distribution of the Berry curvature is shown.
We see the peaks at $p = 0$, which are related to the physical pole at $m=0$.
The peak is negative at $ma=0.5$ and positive at $ma=-0.5$.
This change indicates the topological transition at $m=0$.
The Chern number $N$ is given by the integral of the Berry curvature.
The Chern number of the ground state is obtained by taking imaginary time $\tau$ large enough.
The imaginary-time dependence is shown in Fig.~\ref{figN}.
For $m>0$, the result is independent of $\tau$, and thus trivially $N=0$.
For $m<0$, the result depends on $\tau$.
We look at the region of $\tau \ge 12$ and conclude $N=1$.
These results are consistent with our expectation in Fig.~\ref{figfree}.
The data of non-interacting simulation and quenched Monte Carlo simulation are shown in Fig.~\ref{figN}.
The results are the same in the present parameters, while they can be different by interaction effects in general.

\section{Summary}

The lattice QCD simulation of the Berry curvature was formulated.
The validity of the formulation was successfully confirmed in a simple example, the (2+1)-dimensional Wilson fermion.
The formulation will be applicable to physical phenomena in realistic systems: the chiral magnetic and vortical effects in QCD, the quantum Hall effect and topological insulators in condensed matter physics, etc.

\section*{Acknowledgments}

The author is supported by JSPS KAKENHI Grant Number JP15K17624.
The numerical simulations were carried out on SX-ACE in Osaka University.

\end{document}